\documentclass[%
 reprint,
 superscriptaddress,
showpacs,
 amsmath,amssymb,
 aps,
]{revtex4-1}

\usepackage{graphicx}
\usepackage{dcolumn}
\usepackage{bm}
\usepackage{color}
\usepackage[usenames,dvipsnames]{xcolor}

\begin{document}

\preprint{APS/123-QED}

\title{Attosecond XUV vortices from high-order harmonic generation}

\author{C. Hern\'andez-Garc\'ia}
\email{carloshergar@usal.es}
\affiliation{Grupo de Investigaci\'on en \'Optica Extrema, Universidad de Salamanca, E-37008, Salamanca, Spain}
\author{A. Pic\'on}
\affiliation{Argonne National Laboratory, Argonne, Illinois 60439, USA}
\author{J. San Rom\'an}
\affiliation{Grupo de Investigaci\'on en \'Optica Extrema, Universidad de Salamanca, E-37008, Salamanca, Spain}
\author{L. Plaja}
\affiliation{Grupo de Investigaci\'on en \'Optica Extrema, Universidad de Salamanca, E-37008, Salamanca, Spain}

 \date{\today}

\begin{abstract}
We present a theoretical study of  high-order harmonic generation (HHG) and propagation driven by an infrared field carrying orbital angular momentum (OAM). Our calculations unveil the following relevant phenomena: extreme-ultraviolet harmonic vortices are generated and survive to the propagation effects, vortices transport high-OAM multiple of the corresponding OAM of the driving field and, finally, the different harmonic vortices are emitted with similar divergence. We also show the possibility of combining OAM and HHG phase-locking to produce attosecond pulses with helical pulse structure.

\end{abstract}

\maketitle


Helical phased beams, also called optical vortices, are structures of the electromagnetic field with a spiral phase ramp about a point-phase singularity. The phase wind imprints an orbital angular momentum (OAM) to the beam, in addition to the intrinsic angular momentum associated with the polarization \cite{Allen92,Soskin01,Calvo06}. As light-matter interaction is inherently connected with the exchange of momentum, OAM can also be transferred to atoms \cite{Atoms1,Atoms2,Atoms3}, molecules \cite{Molecules1,Molecules2}, or an ensemble of atoms \cite{Ensemble1,Ensemble2,Ensemble3,Ensemble4,Ensemble5}. Optical vortices have potential technological applications in optical communication \cite{wang12,cai12}, micromanipulation \cite{Grier03}, phase-contrast microscopy \cite {Furhapter05,Jesacher05}, and others \cite{Torres_Libro_2011}. In the recent years an interest has burgeoned in imprinting phase singularities in the XUV/X-ray light generated both in synchrotron and X-ray free-electron laser (XFEL) facilities \cite{Peele02,Sasaki08,Hemsing11}. In this short-wavelength regime one can drastically reduce the diffraction limit, as well as exploit the selectively site-specific excitation, with an important impact in microscopy \cite{Microscopy1,Microscopy2} and spectroscopy \cite{Veenendaal07}.

Phase singularities can also be imprinted to shorter-wavelength light using high-order harmonic generation (HHG), as it was recently demonstrated by Z\"urch and coworkers producing vortices in the extreme ultraviolet (XUV) \cite{zurch2012}. The confluence of OAM and HHG constitutes an extraordinary promising perspective. For instance, the phase twist is not imprinted directly to the short-wavelength radiation, but to the fundamental field. Therefore, it requires a single setup (diffractive mask, for instance) to imprint OAM to the fundamental field, that will be subsequently transferred to the rainbow of harmonic wavelengths by non-linear conversion. On the other hand, high-order harmonics have extraordinary temporal coherence qualities \cite{krausz09,popmintchev10}, a unique feature that has triggered a revolutionary metrology tool for the temporal characterization of ultrafast processes at the atomic scale (both spatially and temporally) \cite{Corkum07,Salieres12}. Two decades ago, it was demonstrated that the higher part of the HHG spectra can be used to synthesize short XUV pulses of attosecond duration \cite{farkas92,christov97,paul01,hentschel01}. It is, therefore, an appealing possibility if such attosecond pulses can be synthesized with OAM. In addition, the new HHG generation schemes, based on the present development of mid-infrared (mid-IR) laser sources, show that keV X-ray radiation can be obtained from the HHG process in a table-top system \cite{Popmintchev12}, as well as that the zeptosecond era could be closer than expected \cite{hergar13A}. Therefore, there is no fundamental obstacle for up-shifting the XUV OAM beams to extremely short temporal structures, or to generate them in the soft X-ray regime.

A surprising finding of the recent  HHG-OAM experiment \cite{zurch2012} is that the topological charge of the harmonic vortices is nearly equal to one, i.e. that of the fundamental field. This is counterintuitive in terms of the present understanding of HHG with intense fields, in which the phase of the harmonics scales roughly with the harmonic order. At present, it is recognized that a detailed theoretical treatment, including propagation effects, is  needed to elucidate some of the fundamental aspects of the HHG OAM conversion \cite{spanner2012}. For instance, to confirm the increase of the topological charge with the harmonic order and, thus, to corroborate that the origin of the lower topological charge of the harmonics detected in \cite{zurch2012} should be attributed to parametric instabilities of the nonlinear propagation and not to the fundamental HHG process. It is presently uncertain that high-charge vortices might be obtained, even if no parametric instabilities are present, as they would have to survive phase-matching effects during propagation. Finally, it is also an open question whether attosecond HHG OAM pulses can be generated and be, also, resilient to propagation.

In this Letter we present a pioneering theoretical study of the HHG process from an IR field carrying OAM of topological charge $\ell$, in the low intensity regime. In this regime, the nonlinear propagation instabilities, which could affect drastically the observations \cite{zurch2012}, can be neglected. We prove that: 1) XUV harmonics vortices are generated and survive to the propagation effects, 2) each harmonic has a topological charge of $q\ell$, $q$ being the harmonic order, 3) all the harmonics are emitted with similar divergence and, 4) attosecond pulses carrying OAM can be generated and also survive propagation.


We compute harmonic propagation using the electromagnetic field propagator \cite{hergar10}. The HHG emission of the single-atom sources is computed using the SFA+ method, an extension of the standard strong field approach with good quantitative accuracy \cite{perez-hdez09A}. Every point of the target (gas cell or gas jet) is treated as source of an elementary wave generated by an ensemble of atoms, which is then propagated to the detector. The final field at the detector is, thus, the coherent addition of these elementary waves. One of the advantages of this method is that is well-fitted to compute high-order harmonic propagation in non-symmetric geometries, therefore, it is specially suited for computing HHG driven by beams carrying OAM. The method has been successfully used for describing regular HHG with near- and mid-IR lasers, in good agreement with experiments \cite{hergar13B,Popmintchev12}. More information on this method and the SFA+ may be found in Refs. \cite{perez-hdez11A,hergar12}. 

In order to implement the OAM beams, we consider the Laguerre-Gaussian (LG) modes in the paraxial approximation \cite{Allen92}, i.e. the beam propagates in a well-defined direction (in our case the $z$-axis) as a plane wave modulated by the slowly-varying transverse amplitude (written in cylindrical coordinates) 
\begin{eqnarray}
&LG_{\ell,p}&(\rho, \phi, z)=  E_0 {W_0 \over W(z)} \left( {\rho \over W(z)} \right)^{|\ell|}  L ^{|\ell|}_p \left[ 2\rho^2 \over W^2(z)\right] \times \nonumber \\ 
& & \exp\left(\! {-{{\rho}^2 \over W^2(z)}}\right)   \exp\left({ik{\rho^2 \over 2R(z)}+i\zeta(z)+i\ell\phi} \right)  \! ,
\label{eq:eq1}
\end{eqnarray}
$W(z)=W_0\sqrt{1+(z/z_0)^2}$ is the beam width, where $W_0$ is the beam waist, $W_0=\sqrt{\lambda z_0/\pi}$ ($z_0$ being the Rayleigh length); $R(z)$ is the wavefront radius of curvature, given by $R(z)=z[1+(z_0/z)^2]$;  $\zeta(z)$ is the Gouy phase, which is given by $\zeta(z)=-(|\ell| +2p +1)\tan^{-1}(z/z_0)$; and, $L ^{|\ell|}_p[x]$ are the associated Laguerre polynomials. The indices $\ell=0,\pm1,\pm2,...$ and $p=0,1,2, ...$ correspond to the topological charge and the number of radial nodes of the mode, respectively.

\begin{figure}[h]
   \begin{center}
 \includegraphics[width=8.8cm]{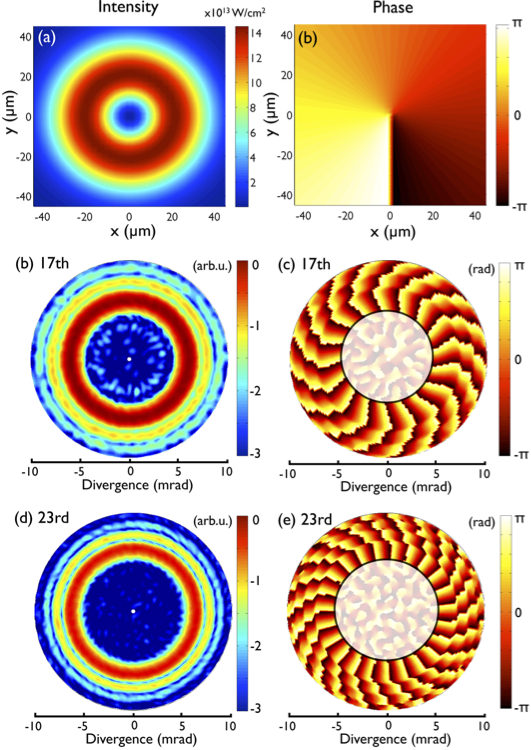}
\caption{(a) Intensity and (b) phase transversal profiles of the $LG_{1,0}$ mode of the fundamental beam at the focus position, with a beam waist of $W_0=30$ $\mu$m. The amplitude $E_0$ in Eq. (\ref{eq:eq1}), is chosen to give a peak intensity of $1.4\times10^{14}$ W/cm$^2$ at the focus. (c,e) Intensity and (d,f) phase angular profiles for the 17th-23rd harmonics. Note that the intensity profiles are given in logarithmic scale. The resulting topological charge can be obtained from (d) and (f) resulting in $\ell=17$ for the 17th and $\ell=23$ for the 23rd harmonic.}
\label{fig:fig1}
  \end{center}
 \end{figure}

We consider as a fundamental field a beam carrying OAM with $\ell=1$, or $\ell=2$, and $p$=0, with a beam waist of $W_0=30$ $\mu$m, and hence a Rayleigh range $z_0$=3.5 mm. The amplitude of the field, $E_0$ is chosen to give a peak intensity at focus of $1.4\times10^{14}$ W/cm$^2$. The laser pulse is assumed to be a $\sin^2$ envelope of 5.8 cycles (15.4 fs) FWHM and 800 nm wavelength. In Figure \ref{fig:fig1}(a,b) we present the intensity and phase profiles of the $LG_{1,0}$ mode at the focus position for the above parameters. As it can be observed in plot (b), the term $e^{i\phi}$ imprints an azimuthal phase variation on the beam from $-\pi$ to $\pi$.

In our simulations, the OAM beam is focused into an argon gas jet, which is directed along the $x$-axis, and modeled by a Gaussian distribution along the $y$ and $z$ dimensions (whose FWHM is 500 $\mu$m), and a constant profile along its axial dimension, $x$, with a peak density of $10^{17}$ atoms/cm$^3$. The gas jet is located after the focus position, because in this region the Gouy phase and the intrinsic phase of the harmonics tend to compensate each other, resulting in optimal longitudinal phase-matching conditions \cite{salieres95A,balcou97}. In addition, as mentioned above, the pulse intensity and the gas density are chosen such that the nonlinear effects of the propagation can be neglected both for the IR as well as the XUV fields. 

\begin{figure}[h]
   \begin{center}
 \includegraphics[width=8.5cm]{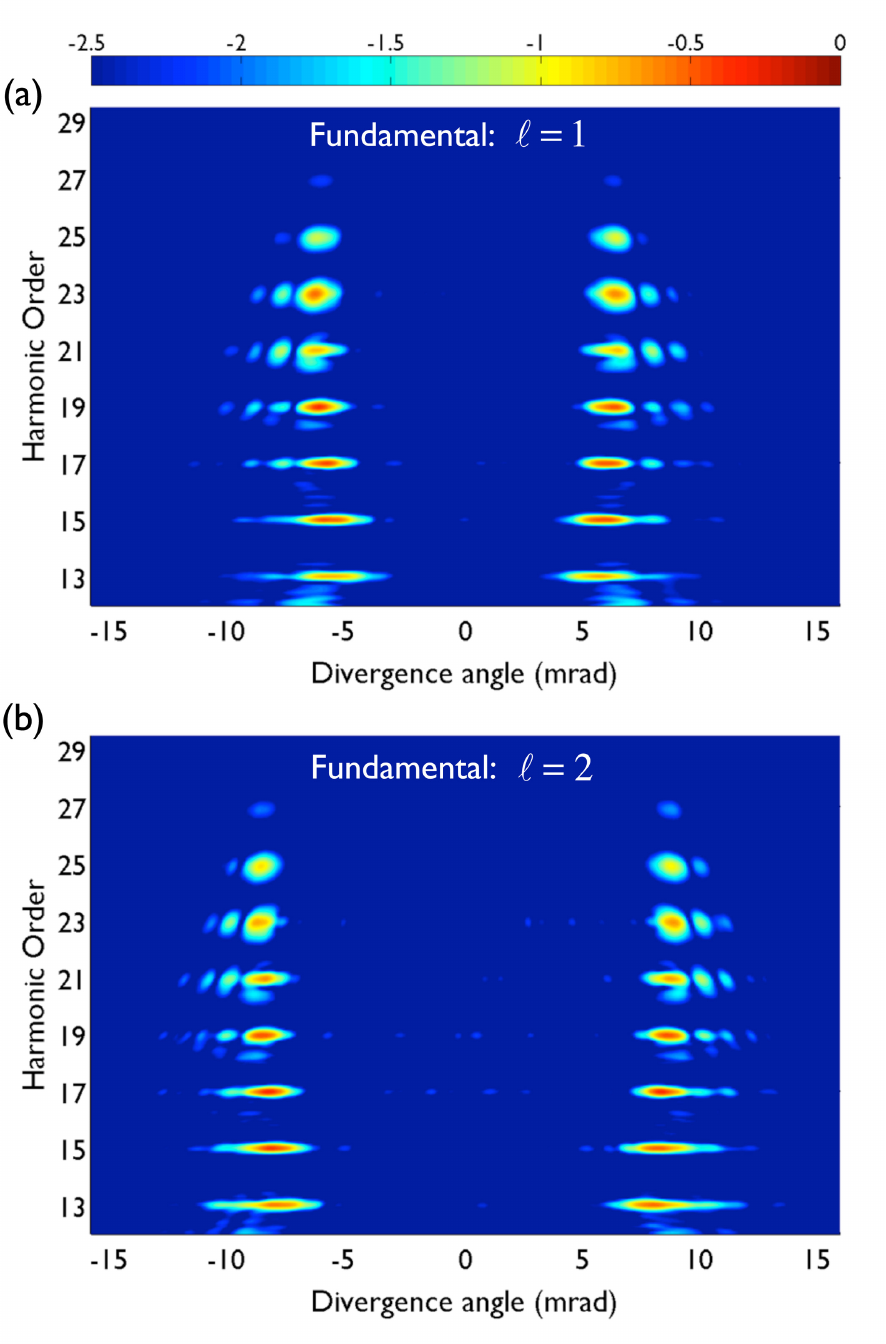}
\caption{Angular distribution of the harmonic spectra generated by an optical vortex of charge (a) $\ell=1$ ($LG_{1,0}$) and (b) $\ell=2$ ($LG_{2,0}$). The parameters for plot (a) are the same as in Fig. \ref{fig:fig1}, whereas for plot (b) we have changed the topological charge, $\ell=2$, and the amplitude $E_0$ in Eq. (\ref{eq:eq1}) has been increased to give a peak intensity at the focus of $1.4\times10^{14}$ W/cm$^2$. The radii of the generated XUV vortex is similar for all the harmonics, being longer for the higher topological charge. The spectra are presented assuming a Al filter plate of 500 nm in thickness.}
\label{fig:fig2}
  \end{center}
 \end{figure}

In Fig. \ref{fig:fig1}(c-f) we present the intensity-phase angular profiles of the 17th (c,d) and 23rd (e,f) harmonics generated with the $LG_{1,0}$ mode. Two main conclusions arise from these plots. First, the radius of the annular intensity distribution is similar for the two selected harmonics, thus, they are emitted with similar divergence. In fact, similar divergence emission can be observed for the whole harmonic spectra --see Fig. \ref{fig:fig2}(a)--. Secondly, the phase-plots show that the charge of the $qth$-order harmonic is $q$, which, as mentioned above, is expected from the HHG theory \cite{spanner2012}. The harmonics generated with the $LG_{2,0}$ mode (not shown) fulfill also these properties, appearing with a similar divergence angle and the $qth$-harmonic having $2q$ topological charge. 

Figure \ref{fig:fig2}(a) shows the angular profile of the spectrum diverging along the y-axis (transverse to propagation) for the same parameters as in Fig. \ref{fig:fig1}. We can observe that the harmonics in the plateau region exhibit a double-fringe profile corresponding to the vortex intensity-distribution, whose radius is similar for all the harmonics, a feature that was also observed in \cite{zurch2012}. For a fundamental beam of topological charge $\ell=2$, in particular a $LG_{2,0}$ mode, we also observe that the high-harmonics are emitted as XUV vortices with si\-mi\-lar radii --see Fig. \ref{fig:fig2}(b)--, being the radii longer than those obtained with $\ell=1$. Therefore, HHG leads to a perfect vortex generation process in terms of its applicability \cite{ostrovsky2013}, as all these XUV vortices of topological charge $q\ell$ are emitted with similar size.

  \begin{figure*}[th]
 \begin{center}
 \includegraphics[width=16.5cm]{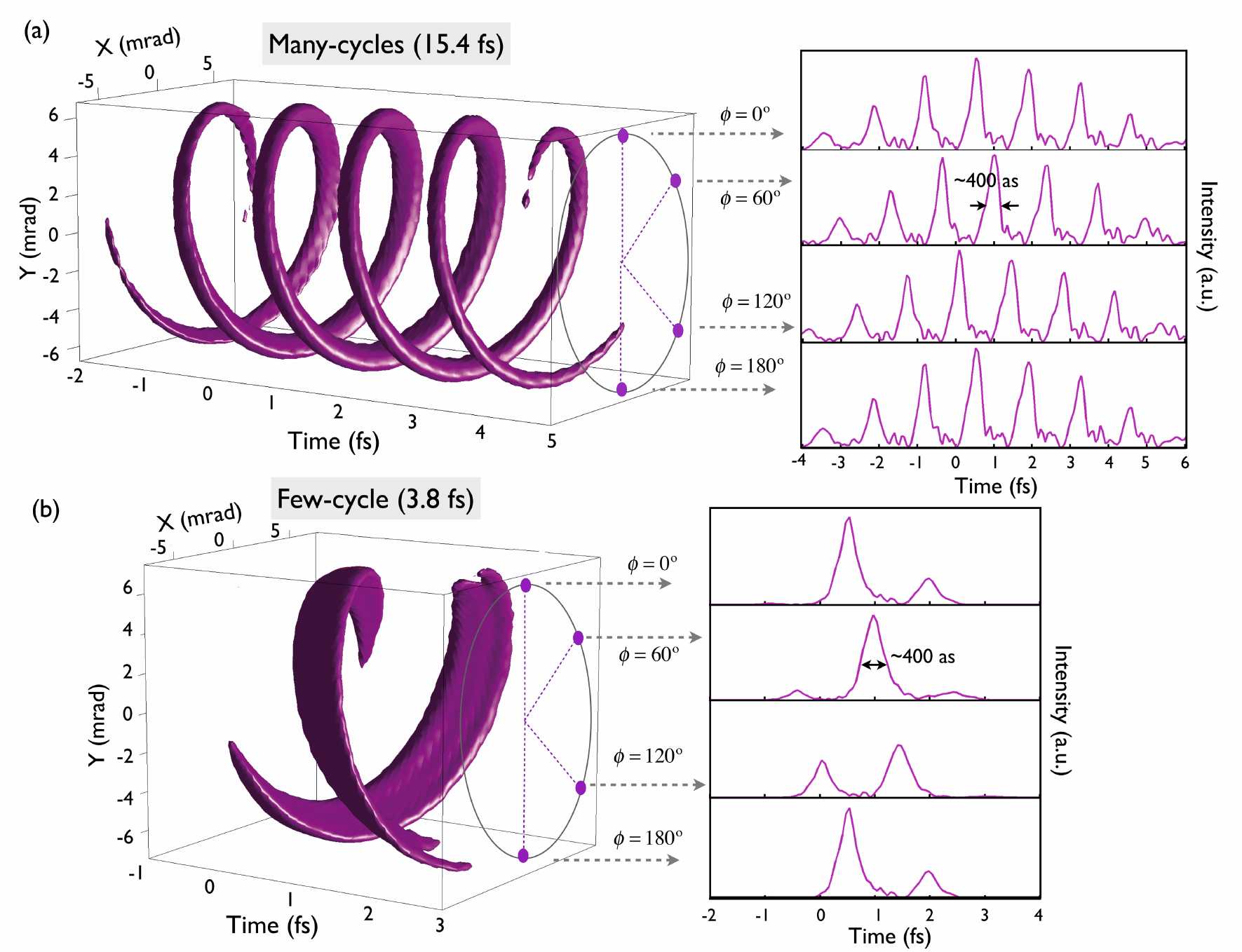}
 \caption{Temporal evolution of the high-harmonic signal driven by (a) a multi-cycle pulse --5.8 cycles (15.4 fs) FWHM-- and (b) a few-cycle pulse --1.4 cycles (3.8 fs) FWHM--. The XUV OAM is emitted in the form of an helical attosecond structure, whose detailed form is presented on the right side for different azimuthal angles $\phi$: 0$^\circ$, 60$^\circ$, 120$^\circ$ and 180$^\circ$. }
\label{fig:fig3}
  \end{center}
 \end{figure*}
 
Let us now look at how the XUV vortices are emitted temporally. One of the most exciting perspectives of high-order harmonic generation by intense lasers is the possibility of synthesizing XUV pulses of sub-femtosecond duration \cite{farkas92}. An attosecond pulse train ($1$ as $=10^{-18}$ s) is obtained by the selection of the higher frequency part of the harmonic spectrum that conform the plateau region \cite{paul01}. For the correct synthesis, the spectrum should approximately satisfy the two following conditions: on one side its structure should approach to that of a frequency comb, in which the harmonic intensities are similar; on the other side, the relative phase between the harmonics should be nearly constant (phase locking) \cite{farkas92}. These two conditions are well satisfied in the typical harmonic spectrum generated during the interaction of an intense field with an atom. Fortunately, HHG driven by OAM beams leads to the generation of XUV OAM beams with similar divergence, therefore allowing for the synthesis of an helical attosecond pulse train. 

In figure \ref{fig:fig3}(a) we present the temporal evolution of the high-harmonic signal shown in Figs. \ref{fig:fig1} and \ref{fig:fig2}(a). As the fundamental laser pulse consists in many cycles --5.8 cycles (15.4 fs)--, an helical attosecond pulse train is obtained, i.e. an attosecond pulse train delayed along the azimuthal coordinate according to the phase variation of the fundamental $LG_{1,0}$ beam. Note that the time se\-pa\-ra\-tion at a fixed angle between the pulses of the helical structure is half the period of the driving field (1.33 fs). On the right side of Fig. \ref{fig:fig3}(a) we show the attosecond pulse train obtained at different azimuthal angles $\phi$. Note that although each high-harmonic OAM beam is generated with $\ell=q$, as we are selecting many harmonic orders (see Fig. \ref{fig:fig2}), the pulse train structure remains similar.
 
It is also possible to obtain an isolated attosecond pulse using few-cycle driving pulses \cite{christov97,hentschel01}, since high-order harmonics are then generated in a single rescattering event. For that purpose we show in figure \ref{fig:fig3}(b) the temporal structure of the XUV OAM beam driven by a few-cycle pulse --1.4 cycles (3.8 fs)--. The OAM imprints a different carrier-envelope phase (CEP) in the fundamental pulse over the azimuthal angle $\phi$, and thus, the helical attosecond structure varies from an isolated attosecond pulse, generated around $\phi=60^\circ$, to a double pulse structure, around $\phi=120^\circ$. As a consequence, we obtain in a single-shot the map of all the attosecond pulse structures for different CEP values. We believe that these helical attosecond structures, obtained in a single-shot, are a powerful tool for pump-probe experiments with attosecond resolution.


In conclusion, we have presented theoretical calculations of the HHG process driven by an intense IR beam carrying OAM with topological charge $\ell$. We have obtained XUV harmonic vortices with topological charge of $q\ell$, $q$ being the harmonic order. In addition, we have shown that all the harmonics are emitted with similar  divergence and that these structures are robust under propagation. We have also proven that the highest order harmonics can be used to synthesize OAM attosecond pulses, that exhibit an attosecond helical structure. We believe that the combination of the properties of optical vortices carrying OAM with the spatio-temporal characteristics of high-order harmonic generation and propagation opens a new perspective in ultrafast science.

\begin{acknowledgments}
We acknowledge I. J. Sola for valuable discussions, and support from Spanish MINECO (FIS2009-09522) and Centro de L\'aseres Pulsados (CLPU). A.P. acknowledges fruitful discussions with I. McNulty and S.H. Southworth, and the financial support of the U.S. Department of Energy, Basic Energy Sciences, Office of Science, under contract \# DE-AC02-06CH11357. 
\end{acknowledgments}


\bibliography{apssamp}

\end{document}